\documentstyle[epsf]{mn}

\title[]{RX J0806+15: the shortest period binary?}

\author[Ramsay, Hakala \& Cropper]{
Gavin Ramsay$^{1}$, Pasi Hakala$^{2}$, Mark Cropper$^{1}$\\
$^{1}$Mullard Space Science Lab, University College London,
Holmbury St. Mary, Dorking, Surrey, RH5 6NT, UK\\
$^{2}$Tuorla Observatory, University of Turku, V\"ais\"al\"antie 20, 
21500 Piikki\"o, Finland}
\date{Received: 2002 Jan 25, Accepted: 2002 Mar 5}

\begin{document}
\outer\def\gtae {$\buildrel {\lower3pt\hbox{$>$}} \over 
{\lower2pt\hbox{$\sim$}} $}
\outer\def\ltae {$\buildrel {\lower3pt\hbox{$<$}} \over 
{\lower2pt\hbox{$\sim$}} $}
\newcommand{\ergscm} {ergs s$^{-1}$ cm$^{-2}$}
\newcommand{\ergss} {ergs s$^{-1}$}
\newcommand{\ergsd} {ergs s$^{-1}$ $d^{2}_{100}$}
\newcommand{\pcmsq} {cm$^{-2}$}
\newcommand{\ros} {\sl ROSAT}
\newcommand{\exo} {\sl EXOSAT}
\def\rchi{{${\chi}_{\nu}^{2}$}}
\newcommand{\Msun} {$M_{\odot}$}
\newcommand{\Mwd} {$M_{wd}$}
\def\Mdot{\hbox{$\dot M$}}
\def\mdot{\hbox{$\dot m$}}

\maketitle

\begin{abstract}

The X-ray source RX J0806+15 was discovered using {\ros}, and shows an
X-ray light curve with a prominent modulation on a period of 321.5
sec.  We present optical observations in which we report the detection
of its optical counterpart. We find an optical period consistent with
the X-ray period. We do not find convincing evidence for a second
period in the data: this implies the 321.5 sec period is the orbital
period. As such it would be the shortest period stellar binary system
yet known. We discuss the nature of this system. We conclude that an
isolated neutron star and an intermediate polar interpretation is
unlikely and that a double degenerate interpretation is the most
likely.

\end{abstract}


\section{Introduction}

Using data taken using the {\ros} satellite, Israel et al (1999) and
Burwitz \& Reinsch (2001) found that RX J0806.3+1527 (hereafter RX
J0806+15) was modulated on a period of 321.5 sec. Further, the
amplitude was 100 per cent with zero X-ray flux for half of the 321.5
sec period. Burwitz \& Reinsch (2001) found a $V\sim$21 mag object
within the X-ray error-box and considered it the most likely optical
counterpart (star `A' in their Figure 3). Using the Digital Palomar
Survey, Israel et al (1999) did not find an optical counterpart in the
$R$ band down to $\sim$20, although they did find a blue object close
to the X-ray position in the Automatic Plate Measuring machine
catalogue (Irwin, Maddox \& McMahon 1994).

Israel et al (1999) suggested that RX J0806+15 was a relatively
distant intermediate polar (IP: a weakly magnetic cataclysmic
variable) or a nearby isolated neutron star. Burwitz \& Reinsch (2001)
ruled out the isolated neutron star option based on its soft X-ray
spectrum. For all known IPs, none show an X-ray light curve which
shows zero flux for half the X-ray period. Indeed, it is difficult
(although not impossible) to envisage a IP geometry which would
produce this type of X-ray light curve (Cropper et al 1998).  For this
reason Burwitz \& Reinsch (2001) suggested that this option was not
likely.

They did however, notice the resemblance of the X-ray folded light
curve with that of another {\ros} discovered source RX J1914+24 which
has a period of 569 sec. Cropper et al (1998) proposed that this
object was a double degenerate polar -- an interacting binary
consisting of two white dwarfs, one of which is magnetic. In this
interpretation, the 569 sec period was the binary orbital period and
also the spin period of the magnetic white dwarf. Recently, Wu et al
(2002) have proposed that RX J1914+24 is driven by electrical power,
while Marsh \& Steeghs (2002) and Ramsay et al (2002) have proposed it
is a double degenerate Algol system.

We have made optical observations of the field of RX J0806+15 with the
goal of identifying the optical counterpart of RX J0806+15 and thereby
making progress in understanding the nature of this system.

\section{Observations}

Optical observations were carried out using the 2.5m Nordic Optical
Telescope on La Palma on the nights of 14/15 and 15/16 Jan 2002. The
instrument was the Andalucia Faint Object Spectrograph and Camera
(ALFOSC), the detector being a Loral 2048x2048 CCD. Conditions were
photometric. The seeing on the first night was moderate
($\sim$1.1--1.8''), but better on the second ($\sim$0.9--1.2''). The
images were bias-subtracted and flat-fielded in the usual way. On the
first night 600 sec exposures were made of the field of RX J0806+15 in
$BVRI$ bands. A series of shorter (windowed) exposures were also made
of the immediate field in $VI$ bands. On the second night white light
observations were obtained of the immediate field with exposures as
low as 20sec, with readout time being a few seconds. Observations of
Landolt standard stars (PG 0220+132B, PG 0231+051A/D, 94394, Landolt
1992) were obtained. We used the mean atmospheric extinction
co-efficients for La Palma.

\section{The optical counterpart}

\begin{figure}
\begin{center}
\setlength{\unitlength}{1cm}
\begin{picture}(8,6)
\put(0,-2){\includegraphics{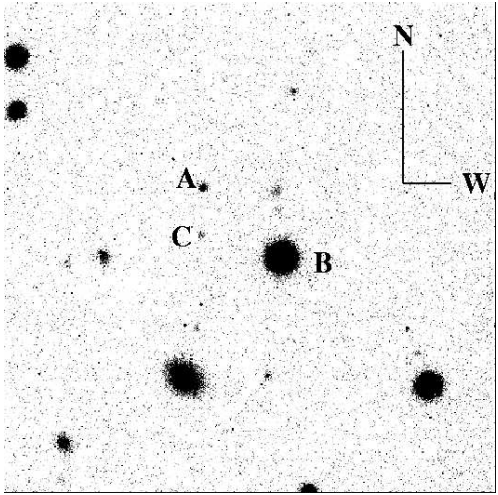}}
\end{picture}
\end{center}
\caption{A $V$ band image of the field of RX J0806+15. Stars A and B refer
to stars mentioned in Burwitz \& Reinsch 2001). Stars A and C are
separated by 11 arc sec.}
\label{chart} 
\end{figure}

We show in Figure \ref{chart} the immediate field of RX J0806+15. The
analysis of our white light observations (\S 4) show that the optical
counterpart of RX J0806+15 is indeed star `A' as suggested by Burwitz
\& Reinsch (2001). The magnitude of star A is: $B=20.9\pm$0.1,
$V=21.1\pm$0.1, $R=21.4\pm0.1$ and $I=21.2\pm$0.1.  It is clearly a
blue object. The Hydrogen column density to the edge of the galaxy in
the direction of RX J0806+15 is relatively low:
$N_{H}=2.7\times10^{20}$ \pcmsq corresponding to $A_{V}$=0.15.

\section{The light curve}

\begin{figure}
\begin{center}
\setlength{\unitlength}{1cm}
\begin{picture}(8,5)
\put(-0.5,0){\includegraphics{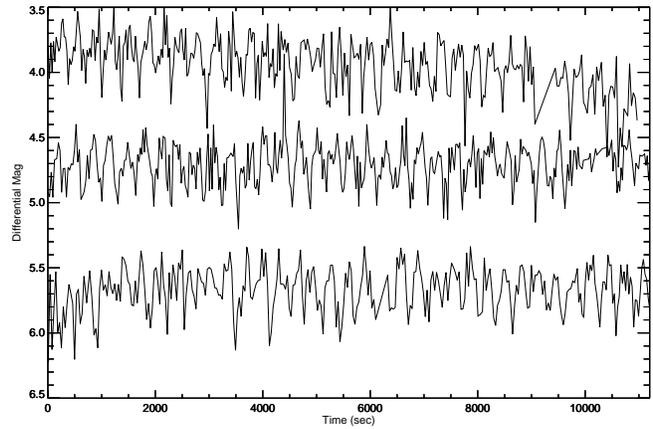}}
\end{picture}
\end{center}
\caption{The unfolded white light light curve obtained on the night
15/16 Jan 2002. The time advances left to right and is then shifted up by 0.9
mag.}
\label{white} 
\end{figure}

On the night 15/16 Jan we made a series of short exposures in white
light: this was a continuous series of observations spanning over
9 hours. The images contained stars `A' and `B' in the chart of
Burwitz \& Reinsch (2001) and also a star $\sim$11'' to the south of
star A - star `C' in Figure \ref{chart}. Our photometry showed that
star B was $V$=15.52 (consistent with that found by Burwitz \&
Reinsch) and $I$=14.64. We performed differential aperture photometry
between stars A and B and also between stars B and C. There was no
significant variation in the differential magnitude between stars B
and C. There was, however, a significant variation between star A and
star B. We show the resulting light curve in Figure \ref{white}. We
also obtained light curves for star A in $VI$ bands on the first
night.

\begin{figure}
\begin{center}
\setlength{\unitlength}{1cm}
\begin{picture}(8,5)
\put(-2,-7){\includegraphics{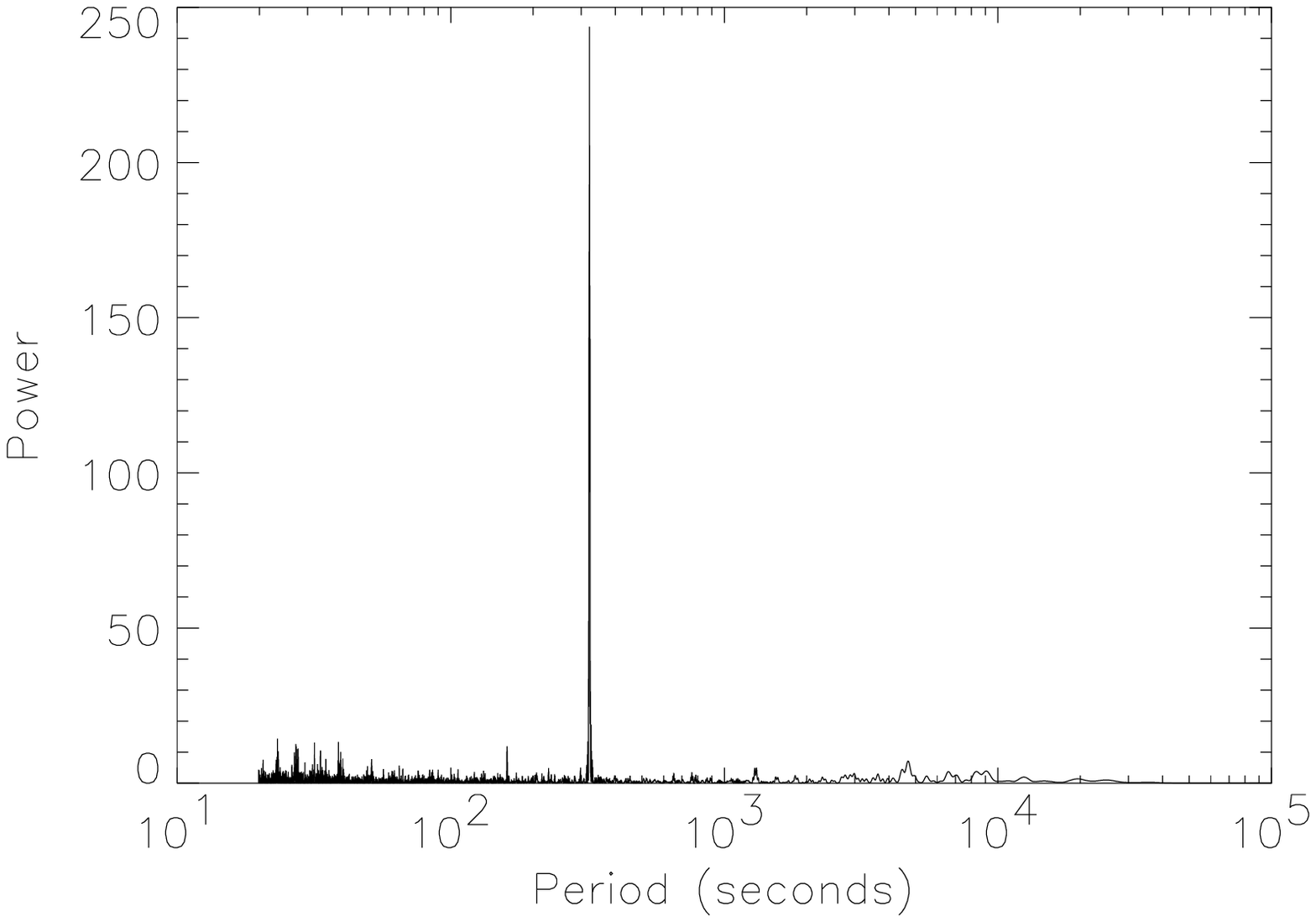}}
\put(2.5,-1.6){\includegraphics{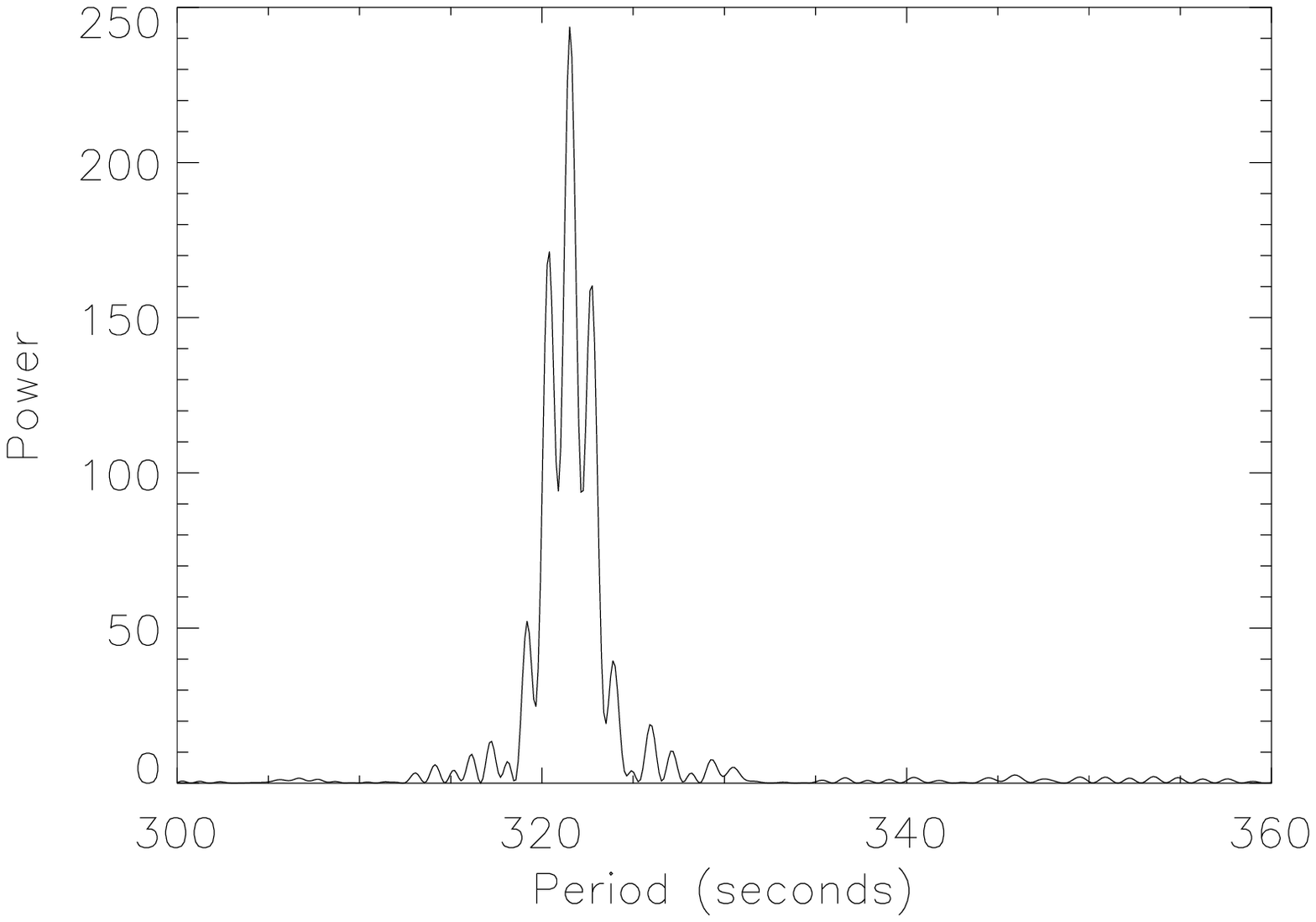}}
\end{picture}
\end{center}
\caption{The power spectrum of the combined white light and $V$ band
data. The peak at 160 sec is the second harmonic of the main peak at
321.5 sec.}
\label{power} 
\end{figure}

Using the white light data we obtained a power spectrum using the
Lomb-Scargle algorithm: a prominent peak is seen at 321.4 sec. We
conclude that star A is the optical counterpart of RX J0806+15. To
refine the period we use an inversion technique based on Bayesian
statistics (Karttunen \& Muinonen 1991). This searches a range of
periods and assumes that the pulse shape can be modelled using a
Fourier expansion of n degrees (we assume n=2). Using this technique
we find a period of 321.404$\pm$0.044 sec. If we include the $V$ band
data (covering nearly 8 hours of data taken on Jan 14/15) we determine
a period of 321.544$\pm$0.014 sec. The power spectra of this combined
dataset is shown in Figure \ref{power}. This is consistent with the
X-ray period of 321.5393$\pm$0.0004 sec (or its alias period of
321.5465$\pm$0.0004 sec) (Burwitz \& Reinsch 2001). Unfortunately,
because of the uncertainty in the periods, we cannot co-phase the
optical and X-ray data (the X-ray data were taken in 1994 and 1995).
We also show in Figure \ref{power} a close-up of the power spectrum
between 300 and 360 sec. The spectrum is remarkably clean and free
from peaks other than those which can be attributed to the window
function.

We folded the white light, $V$ and $I$ band data on the 321.544 sec
period and show them in Figure \ref{fold}. The white light folded
light curve shows an amplitude of 0.30$\pm$0.01 mag (using a second
order Fourier fit to the folded light curve) and shows an asymmetric
light curve with the descent from maximum being more rapid than the
rise to maximum. The $V$ band folded light curve is similar to the
white light curve (amplitude 0.31$\pm$0.03 mag). The signal to noise
of the $I$ band light curve is lower since RX J0806+15 is a relatively
blue object. However, there is some evidence that the amplitude of
the modulation is larger in $I$ than in $V$: 0.54$\pm$0.12 mag.

\begin{figure}
\begin{center}
\setlength{\unitlength}{1cm}
\begin{picture}(8,12.5)
\put(-1.3,5){\includegraphics{fold-wl-cap.ps}}
\put(-1.,10.2){\includegraphics{fold-v-cap.ps}}
\put(-1.7,14.5){\includegraphics{fold-i-cap.ps}}
\end{picture}
\end{center}
\caption{From the bottom: The white light data folded on the 321.528
sec 
period, middle: the $V$ band data and top: the $I$ band data.}
\label{fold} 
\end{figure}

\begin{figure}
\begin{center}
\setlength{\unitlength}{1cm}
\begin{picture}(8,5.5)
\put(-1.5,-6.8){\includegraphics{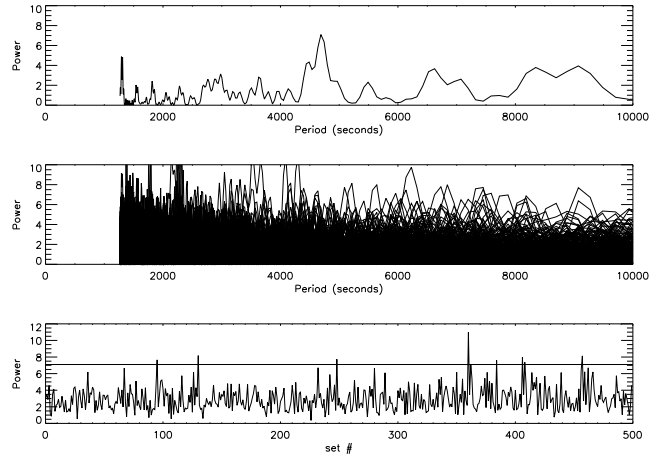}}
\end{picture}
\end{center}
\caption{Top panel: the power spectra of the combined white light plus
$V$ band data in the range 1200--10000 sec. Middle panel: the
resulting power series where we have randomly assigned the data points
to the time points. This has been carried out for 500 trials. Bottom
panel: the power of the highest peak lying in the range 4000--6000 sec
for each of these 500 trials.}
\label{synth} 
\end{figure}

We also searched for longer period modulations in our data. We show in
the top panel of Figure \ref{synth} the power spectrum extending from
1200 sec up to periods of 10000 sec (=2.8hrs) for the white light plus
$V$ band data. (For periods longer than 10000 sec the power is
negligible). This shows a number of peaks, the most prominent being
near 4700sec. To make an assessment of significance of these peaks, we
show in the middle panel of Figure \ref{synth} the power spectrum of
the same data set if we randomly reassign the times of the data points
using 500 trials. In the lower panel of Figure \ref{synth} we show the
power of the highest peak lying in the range 4000-6000 sec for each of
these trials. We find that in $\sim$10 of these 500 trials the maximum
peaks exceeds or is very close to the actual power determined in the
combined white light plus $V$ band light curve. Based on this test we
find the peak at 4700 sec has a significance of only between 2 and
3$\sigma$. We also investigated this further by creating a random
light curve by assigning Gaussian random numbers to the time points of
the combined white light and $V$ band light curve and determining its
power spectrum. Again we conclude that although a period close to 4700
sec maybe present in the data, we cannot be certain.

\section{The location of the optical emission site}

From the colours determined in \S 3, RX J0806+15 is clearly a blue
object. However, it is more difficult to reconcile a blue object with
having a modulation amplitude greater in the $I$ band compared to the
$V$ band. This contrasts with the double degenerate polar RX J1914+24
which shows a larger variation in $V$ compared to $I$. There are
several possible solutions. In the first, we assume that the optical
emission originates on the secondary star and is due to irradiation
from X-rays from the primary. If RX J0806+15 was an accreting double
degenerate polar, the primary white dwarf would be more massive, and
hence smaller in size than the secondary white dwarf. If the heated
face of the secondary covered a small fraction of the area, then the
variation in the cooler area (larger fraction) of the secondary could
be greater than the hotter component.

It is also possible that the X-ray and optical data originate from the
same site on the primary. In this case the higher amplitude of the $I$
band data could be due to the primary having a magnetic field of
sufficient strength to produce cyclotron radiation. A field strength
of $\sim$10MG would emit most strongly at red wavelengths and imply
that the spin period of the primary white dwarf was closely
synchronised with the binary orbital period. Since we cannot co-phase
the optical and X-ray data we cannot distinguish between these
scenarios.

\section{Discussion}
\label{discuss}

We have detected the optical counterpart of RX J0806+15. It has a
period (321.5 sec) consistent with that detected in X-rays. There is
marginal evidence for a second period at $\sim$4700 sec: this needs to
be tested by a longer series of observations. We now consider the
nature of this object.


Burwitz \& Reinsch (2001) claimed that RX J0806+15 was unlikely to be
a isolated neutron star based on the fact that its X-ray spectrum is
very soft. However, they do not show the spectrum, or details of any
fits to that spectrum. The softness ratio of RX J0806+15,
HR1=--0.67$\pm$0.1, is very similar to that of the isolated neutron
star RX J0720--31 ({\ros} public archive). We cannot exclude RX
J0806+15 being an isolated neutron star based on its X-ray softness
ratio. Israel et al (1999) conclude that were RX J0806+15 to be an
isolated neutron star then it would have to be $\sim$10pc distant,
based on a blue magnitude of $\sim$20.5.

If RX J0806+15 was indeed a neutron star at a distance of 10 pc, then
we may expect to detect a significant proper motion. To test this, we
determined the position of RX J0806+15 using our 600 sec $V$ band
image using {\tt ASTROM} (Wallace 1998) (and determining the positions
of stars in the immediate field from the USNO A2 catalogue and the APM
catalogue) and compared it to the position reported by Burwitz \&
Reinsch (2001) and the position of the blue star recorded in the APM
catalogue (Irwin, Maddox \& McMahon 1994) (Table \ref{proper}).  We
find that the position determined by us and the APM catalogue agree to
within 0.5 arc sec over a time interval of 50 years. In contrast, the
isolated neutron star RX J1856-37 has been observed to show a proper
motion of 0.33 arc/yr at a distance of 61pc (Walter 2001). At a
distance of 10pc it would be expected to show a proper motion of 11.9
arc sec/yr. If RX J0806+15 were an isolated neutron star it would have
to have a proper motion very much smaller than that of RX J1856-37. We
consider it unlikely that RX J0806+15 is an isolated neutron star.

\begin{table}
\begin{tabular}{llrr}
\hline 
Date & RA (2000) & Dec (2000) & Reference\\ 
\hline
2002.0 & 08 06 22.98 & +15 27 31.5& 1\\ 
1998.3 & 08 06 22.9 & +15 27 30.2& 2\\ 
1951.1 & 08 06 22.94 & +15 27 31.1& 3\\
\hline
\end{tabular}
\caption{The reported positions of RX J0806+15. References: (1) this
paper, (2) Burwitz \& Reinsch (2001), (3) APM catalogue Irwin, Maddox
\& McMahon (1994).}
\label{proper}
\end{table}

In the intermediate polar (IP) model, the 321.5 sec period would be
associated with the spin period of the accreting white dwarf and the
4700 sec period the binary orbital period. The X-ray softness ratio
(which implies a soft X-ray spectrum) would associate it with the soft
IPs (like PQ Gem and UU Col) which have similar ratios. However, its
X-ray light curve is very atypical for that of an IP. Indeed, Cropper
et al (1998) discussed the possibility that RX J1914+24 was an
IP. They concluded that an IP interpretation was possible if the
binary inclination was close to 90$^{\circ}$ and heavy (phase
dependent) absorption was present, or the soft X-rays were strongly
beamed by some (as yet unknown) mechanism. Further, the light curve
(Figure 2) shows no sign of flickering or flaring activity as is seen
in other IPs. In the PQ Gem the amplitude on the spin period is
$\sim$0.1 and 0.2 mag in $V$ and $I$ bands respectively. This
contrasts with 0.3 and 0.5--0.6 mag in RX J0806+15. We also note the
very clean power spectra (Figure \ref{power}): IPs show complex power
spectra, typically showing a beat period which is not seen in RX
J0806+15. For a spin period of 321.5 sec and an orbital period of 4700
sec, we would expect to observe a beat period around 345 sec: this is
not seen (Figure \ref{power}). Further, no ellipsoidal modulation due
to a main sequence secondary is seen and its colour is too blue. We
consider the IP model unlikely.


Burwitz \& Reinsch (2001) made the suggestion that RX J0806+15 is a
double degenerate polar: a close binary system consisting of a
magnetic white dwarf which is accreting material from a secondary
white dwarf. Indeed, the similarities between RX J0806+15 and RX
J1914+24 (which was proposed as the first double degenerate polar by
Cropper et al 1998) is striking. The X-ray light curves are very
similar: RX J0806+15 has an X-ray period of 321.5 sec, RX J1914+24 569
sec. Both show folded X-ray light curves which have zero flux for
approximately half their period and are asymmetric with the rise to
maximum being more rapid than their descent. The optical and X-ray
light curves of RX J1914+24 are anti-phased, but the uncertainty in
the period of RX J0806+15 prevents us from co-phasing these light
curves.  Although there have been several other suggestions as to the
nature of RX J1914+24 (the electric star model of Wu et al (2002) and
a double degenerate Algol system (Marsh \& Steeghs 2002 and Ramsay et
al 2002) they all assume a double degenerate system. Such a system
would not show a second period lasting several 1000 sec. In the
absence of a convincing second period, we consider the double
degenerate model the most likely scenario.


\section{Summary}

In summary, we consider the isolated neutron star scenario to be
unlikely. We cannot rule out an IP model if the geometry and the
accretion stream absorption was very specific. Since we cannot be
certain that there is a second period on a timescale of 4000-5000 sec
we cannot at this stage rule out a double degenerate model. The nature
of RX J0806+15 therefore, remains open. However, we conclude that the
321.5 sec period marks the rotation period of a white dwarf. If
accretion onto the white dwarf is taking place, then the accretion
flow appears to show little variation.

\section{Acknowledgments}

The data presented here have been taken using ALFOSC, which is owned
by the Instituto de Astrofisica de Andalucia (IAA) and operated at the
Nordic Optical Telescope (NOT) under agreement between IAA and the
NBIfAFG of the Astronomical Observatory of Copenhagen. We also thank
the staff of the NOT for their support. PJH is an Academy of Finland
research fellow.

{\it Footnote: After this paper was submitted, Israel et al (2002)
published a report in an IAU Circular announcing the discovery of
optical modulation in star A of Burwitz \& Reinsch (2001).}

\end{document}